\documentclass{aa}  

\usepackage{graphicx}
\usepackage{amsmath}
\usepackage{color}
\usepackage{comment}
\usepackage{txfonts}
\begin{document}

   \title{Impact of dust diffusion on the rim shape of protoplanetary disks}


   \author{B. N. Schobert
          \inst{1}
          \and
          A. G. Peeters\inst{1}
          }

   \institute{Department of Astrophysics, University of Bayreuth,
              Universit\"atsstraße 30, Bayreuth, Germany\\
              \email{benjamin.schobert@uni-bayreuth.de}
             }

   \date{Received \today}

 
  \abstract
   {Multiple mechanisms are known to give rise to turbulence in protoplanetary disks, which facilitates the accretion onto the central star. Small dust particles that are well coupled to the gas undergo diffusion due to this turbulent motion.}
   {This paper investigates the influence of turbulence induced dust diffusion on the equilibrium of protoplanetary disks.}
   {The model accounts for dust sublimation, radiative transfer with the flux-limited diffusion approximation and dust diffusion. It predicts the density and temperature profiles as well as the dust-to-gas ratio of the disk.}
   {It is shown that dust diffusion can have a large impact: assuming the dust survives for $10^4$ seconds or longer before it can be evaporated, leads the dust diffusion to widen the inner disk considerably. The latter effect is generated through a feedback mechanism as the diffusion length is much smaller than the disk width. With increasing dust diffusion, the inclination of the inner rim towards the stellar radiation becomes steeper until it is almost vertical. The temperature range of evaporation and condensation, which is linked to the dust composition, has no influence on this effect.}
   {For realistic parameters dust diffusion can not be neglected when determining the equilibrium of the disk. Stronger turbulence inside the disk induces more dust diffusion. Therefore, the dust density grows more gradually over a greater distance and less radiation reaches the disk surface. The new equilibrium shape of the disk is more inclined towards the star. This effect is universal and independent of the specific dust composition.}


   \maketitle
%

\section{Introduction}

Herbig Ae Be stars are intermediate mass stars younger than the Sun. They have accretion disks that are believed to be the origin of planets. In  recent years the resolution in which protoplanetary disks (PPDs) can be observed has increased making the inner rim more visible and allowing for constraints on its structure. \cite{Laza17} found the ratio between disk height and radius in the inner disk to be $z/R \approx 0.2$, which is larger than predicted by previous theoretical models \citep{Vin07, Mul12} by a factor of approximately two. The shape of the inner rim has been debated for a long time, because it determines the physical conditions in the dusty planet-forming region.

In an early development \cite{DDN} modeled the inner rim as a cylindrical wall that truncates the disk where the evaporation temperature of the dust is reached. \cite{IN} took into consideration that this evaporation temperature is dependent on the gas density, which leads to a rounded rim. \cite{FL16} then incorporated this function for the evaporation temperature into their radiative transfer model confirming the structure of the rim. An alternative explanation for a rounded rim which does not rely on a density dependent $T_\textrm{ev}$ is the different speeds of dust settling for two species of dust particles \citep{Tannir07}.

Especially the transition between no dust and dust at the inner rim can be difficult to resolve, because of the high temperature gradient and the significant change in opacity. In \citep{FL16} the dust-to-gas ratio was chosen such, that not too much star light would be absorbed within a single cell to ensure a smooth transition. Using the same dust description \citep{Schobert19} found model solutions with interesting feedback loops when considering the absorption of star light. That work found temperature deviations manifesting into dust walls traveling through the disk. As fluctuations in the spectral energy distribution (SED) with a period of days to weeks are a common observation in PPDs \citep{Fla14}, exploring these fluctuations could prove to be insightful. This paper continues to investigate the idea that these were related to the treatment of dust at its evaporation temperature. In light of the observed feedback and to describe the transition at the dust boundary more realistically this work introduces the effect of diffusion into the description of the dust.

A complete description of the inner rim necessarily includes dust diffusion. Gas accretion is commonly believed to be facilitated by turbulence. This turbulence will inevitably induce diffusion into the dust particles as long as they are strongly bound to the gas. This builds upon the notion that for micrometer dust particles the Stokes number is close to one and therefore the dust diffusion coefficient will be nearly equivalent to the gas diffusion coefficient \citep{DULL08}. 
Since both dust formation and destruction are not instantaneous processes, dust will be found slightly outside of its equilibrium position in regard to temperature.
The timescale in which dust can be formed or destroyed is of order $10^5 - 10^7$ seconds \citep{MOR88, TA11, NA07, Lenz95}. Dust diffusion will therefore take place on that timescale.
This addition is also beneficial because it naturally smooths the transition from no dust to dust, which was previously difficult to resolve. Therefore, this paper presents a method of incorporating the diffusive motion of the dust particles into the model of \citep{FL16, Schobert19}. 

A short back of the envelope calculation of the length scale corresponding with dust diffusion $l_\textrm{diff}= 6.5 \cdot 10^{-4}\,$AU (for details see sec. \ref{sec:diffTime}) would suggest that it is negligible to the overall conformation of the inner rim. However, it will be shown in this paper that the impact of dust diffusion is much larger than can be expected from the estimate given above.
   
This paper is structured as follows: Section \ref{sec:modEq} explains the model equations introduced to describe the dust diffusion. In Section \ref{sec:numIm} the numerical implementation and convergence criteria are briefly detailed. Section \ref{sec:results} outlines the results and qualitative changes compared to a model without dust diffusion. Section \ref{sec:discussion} discusses the results in comparison to an analytical prediction and with respect to the rim shape. The paper concludes in section \ref{sec:concl} with a summary of the results.

\section{Model equations}
\label{sec:modEq}
The model equations used in this work for radiative transfer and hydrostatic equilibrium follow those proposed in \citep{FL16, Schobert19}. The exact equations and numerical methods are detailed in \citep{Schobert19}. For brevity only the changes applied to the model compared to \citep{Schobert19} are outlined in the following chapter.
\subsection{Dust-to-gas ratio}
\label{sec:d2g}

Dust sublimation is of deciding importance for the configuration of the inner rim and thereby for the evolution of the complete disk. Because the dust blocks irradiation of the area behind it and is heated by the star and infrared radiation emitted by the dust itself, it greatly affects the temperature profile of the disk. The transition distance between vapor and condensed dust is very thin and requires special attention.

In order to resolve this thin layer the dust sublimation formula from \cite{FL16} smooths the transition over a temperature range of 100 K and uses the tangens hyperbolicus as a model function. The formula is
\begin{equation}
	f_{\mathrm{d2g,old}} =\begin{cases}
	
	  \frac{f_{\Delta \tau}}{2} \left\lbrace 1- \tanh \left[\left(\frac{T-T_{\mathrm{ev}}}{100\, \textrm{K}}\right)^3\right] \right\rbrace \cdot

	 \left\lbrace \frac{1- \tanh(1-\tau_*)}{2} \right\rbrace, & \text{if $T>T_{\text{ev}}$}\\

 \frac{f_0}{2} \left\lbrace 1- \tanh (20-\tau_*)\right\rbrace + f_{\Delta \tau}, & \textrm{otherwise,}
\end{cases}
\label{d2g}
\end{equation}
with the dust evaporation temperature $T_{\mathrm{ev}}$, the reference dust-to-gas ratio $f_0$ and the transition dust-to-gas ratio $f_{\Delta \tau}$.
For the dust evaporation temperature the fitting model proposed by \cite{IN}
\begin{equation}
T_{\mathrm{ev}} = 2000 \, \textrm{K} \, \left(\frac{\rho}{1 \, \textrm{g\,cm}^{-3}}\right)^{0.0195}
\end{equation}
is used. It describes the dependence of the evaporation temperature on the gas density for silicate grains.
The transition dust-to-gas ratio $f_{\Delta \tau}$ is defined as
\begin{equation}
f_{\Delta \tau} = \frac{\Delta \tau_*}{\rho_{\mathrm{gas}} \kappa_{\mathrm{dust}}(\nu_*) \Delta r}= \frac{0.3}{\rho_{\mathrm{gas}} \kappa_{\mathrm{dust}}(\nu_*) \Delta r}
\end{equation}
with $\Delta r$ being the radial size of one grid cell. The transition optical depth of $\Delta \tau_* = 0.3$ is chosen so that the absorption of the radiation at the rim can be resolved \citep{FL16}. This numerical remedy ensures that the transition can be described within each chosen radial resolution, it is however not necessitated by the physics involved.
Furthermore, it is useful for numeric stability to impose a minimum value of $f_{\mathrm{d2g}}^{\mathrm{min}}= 10^{-10}$. The maximum value of $f_0 = 10^{-2}$ is chosen, because it reflects the amount of dust present in the interstellar medium \citep{LiDrain} and therefore represents the maximum ratio in the protoplanetary disk.

This method however becomes discontinuous if an optically thick area is heated above evaporation temperature. This was observed in \cite{Schobert19} in the form of dust waves traveling through the inner hole of the disk. To remove this discontinuity from the description a new formula for the dust-to-gas ratio is presented in this paper, that is simpler and continuous:

\begin{align}
	f_\mathrm{d2g,new} &=  \frac{f_0}{2} \left\lbrace 1- \tanh \left[\left(\frac{T-T_{\text{ev}}}{\Delta T_\textrm{dust}}\right)^3\right] \right\rbrace \left\lbrace \frac{1- \tanh(1-\tau_*)}{2} \right\rbrace,
	\label{eq:fd2gnew}
\end{align}
where $f_{\Delta \tau}$ was replaced by $f_0$. This also means that no separation between above and below evaporation temperature is necessary anymore, because the second case of eq. \eqref{d2g} is incorporated into the first. It is also preferable to get the same result ($f_\textrm{d2g}= f_0$ for low temperatures and high optical depths) as previously with just one equation that is continuous. Additionally a new variable $\Delta T_\textrm{dust}$ is introduced to describe the size of the temperature range over which the dust-to-gas transition occurs. This parameter is dependent on the composition of the dust and in this paper generally $\Delta T_\textrm{dust} = 100\,$K is chosen, but different values are investigated. This allows the model to be more general and section \ref{sec:delT} details the impact of different temperature ranges on the results.

The use of this new formula in the numerical implementation becomes possible by overcoming a pitfall of the previous model, where all of the radiation would be absorbed in a single cell, if the dust-to-gas ratio was allowed to rise too rapidly. To ensure a smooth transition the new model now relies on the effect of dust diffusion instead of a staggered dust-to-gas ratio. How dust diffusion was incorporated into the model is explained in the following section.

\subsection{Dust diffusion}
\label{sec:diff}

To derive a description for dust diffusion one starts with the continuity equation and inserts the advective and diffusive flux components:

\begin{align}
	\frac{\partial \rho}{\partial t} &=\nabla \cdot \left( D \nabla \rho\right)-\mathbf {\nabla } \cdot (\mathbf {v} \rho)+R,
\end{align}
where $\rho$ is the density, $D$ the diffusion coefficient, $\mathbf{v}$ the fluid velocity and R any source or sink terms necessary for dust description. Because of the relative smallness of the diffusion term one can neglect it at first and solve the equations for the purely advective case. This leads to the hydrostatic solution for the gas density \citep{FL16, Schobert19} and to $\rho_{\mathrm{dust}} = f_\textrm{d2g} \rho_{\mathrm{gas}} = \rho_0$ for the dust density as a best approximation considering that $R$ is not fully known. Inserting this lowest order solution $\rho_0$ into the first order terms leads to:
\begin{align}
	\frac{\rho_\textrm{diff} - \rho_0}{\tau_\textrm{diff}} &=\nabla \cdot \left( D \nabla \rho_\textrm{diff}\right) \\
	\frac{\rho_\textrm{diff} - \rho_0}{\tau_\textrm{diff}} &=D \, \nabla^2 \rho_\textrm{diff} \label{eq:step2}\\
	\left[1- D \tau_\mathrm{diff} \nabla^2\right]\rho_\mathrm{diff} &= \rho_0,
\end{align}
where $\rho_0$ is the lowest order solution for the density including only advection and $\rho_\textrm{diff}$ is the first order solution including diffusion. A forward differentiation was used together with a typical time $\tau_\textrm{diff}$ to approximate the derivative. Step \eqref{eq:step2} uses that the gradient length of the mass density is expected to be much smaller than the length scale on which the diffusion coefficient $D$ changes. To implement the diffusion the following two implicit equations are computed via the BiCGStab solver:

\begin{align}
\left[1- D \tau_\mathrm{diff} \nabla^2\right]\rho_\mathrm{gas, diff} = \rho_\mathrm{gas} \label{eq:gasdiff}\\
\left[1- D \tau_\mathrm{diff} \nabla^2\right]\rho_\mathrm{dust, diff} = \rho_\mathrm{dust}
\label{eq:dustdiff}
\end{align}
with the dust-to-gas ratio being $f_\mathrm{d2g} = \rho_\mathrm{dust, diff} / \rho_\mathrm{gas, diff}$, $\rho_\mathrm{diff}$ is the respective density of dust or gas after one diffusion time step has taken place, $\tau_\mathrm{diff}$ is the typical time it takes for the dust to condensate or evaporate and $D$ is the diffusion constant of the dust. The idea is that dust particles that are in the process of forming or being destroyed will still be displaced during that time through the intrinsic turbulence or the gas with which they are very well coupled. One can imagine this as a memory of the dust because its destruction and formation is not instantaneous.

The diffusion constant of the dust is connected to the gas diffusion by the Stokes number $St$ \citep{DULL08}:

\begin{align}
D_\textrm{dust} = \frac{D_\textrm{gas}}{1+St} \approx D_\textrm{gas}
\end{align}
and since $St < 1$ for small particles both coefficients approximately match each other \citep{DULL08} and thus $D_\textrm{dust} = \nu = \alpha c_s H$.

The typical condensation time $\tau_\mathrm{diff}$ ranges from $10^5$ to $10^7$ s \citep{MOR88}. The formation timescale of chondrules is $10^5$ s \citep{TA11}. The formation timescales of corundum and hibonite range from $10^5$ to $10^7$ s \citep{NA07}. Forsterite has been observed to nucleate within $1.4 \cdot 10^5 \, \text{s} - 3.3 \cdot 10^5 $ s \citep{Tach14}. Destruction of silicate grains occurs via sublimation and proceeds under quasi-equilibrium conditions \citep{Lenz95,Duschl96}. This implies that the evaporation of silicate dust takes a comparable time to its formation. 
It is a useful simplification to assume one diffusion time as the average time of formation, since the exact time nucleation of dust particles takes and also their evaporation is dependent on the chemistry of the dust composition and also the temperature difference between the evaporation temperature and the environment of the particle. In order to still cover a wide range of possible dust compositions a study for the effect of different formation time scales is performed in section \ref{sec:diffTime}.

\section{Numerical Implementation}
\label{sec:numIm}

This section details the numerical implementation of the above described model as well as the convergence criteria.
The first step is to determine the density distribution by solving the hydrostatic equilibrium equations for a given temperature profile using the method detailed in \citep{Schobert19}. The second step is to implicitly solve the two coupled equations for the temperature and radiation energy density. This is done using the BiCSTAB solver first presented by \cite{VDV}. For preconditioning the incomplete LU-factorization is used. The final steps are calculating the dust-to-gas ratio with formula \eqref{eq:fd2gnew} and then executing one diffusive step. The numerical implementation of the dust diffusion is consistent with the calculation of the radiative equilibrium. It is calculated implicitly for gas and dust density respectively using equations \eqref{eq:gasdiff} and \eqref{eq:dustdiff}. This is done using the BiCGStab solver and LU-factorization as well. After these three steps, the process is repeated starting again with the hydrostatic equilibrium using the newly obtained temperature distribution. The iterations are continued until the convergence criterion is reached.

\subsection{Convergence criterion}
The convergence criterion used is the same as in \citep{Schobert19}, where the relative local change in density and temperature is observed. These normally show an exponential decline in time and converge faster close to the star than further out because of the dynamical time scale. Once the outer regions have completed several orbits and the disk is vertically isothermal, the inner regions have completed many orbits more and should also be converged. However some cases were observed where the inner regions showed periodic behavior. This is discussed in detail in section \ref{sec:waves} and is also consistent with findings from \citep{Schobert19}. In these cases the disk is considered converged if the outer regions fulfill the convergence criterion and the inner periodic region has completed more than 100 cycles.

\subsection{Resolution study}

\begin{figure}
	\centering
	\input{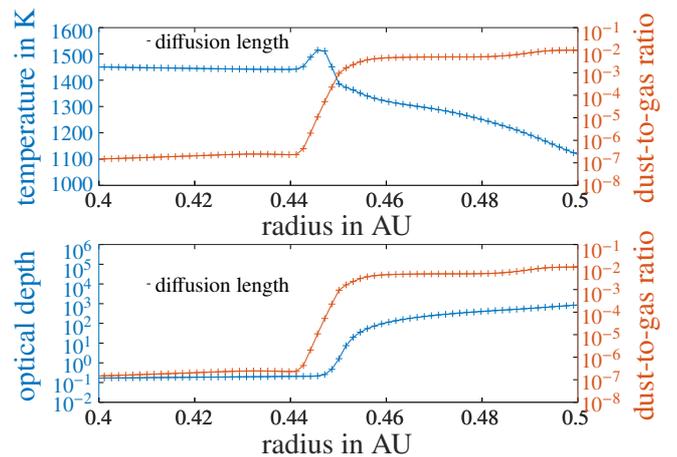}
	\caption{Midplane profiles for temperature, optical depth and dust-to-gas ratio. The crosses represent grid points and the lines are a guide to the eye. The small black line labeled 'diffusion length' represents one diffusion length, which is smaller than one radial grid cell.}%
	
	\label{pic:resolution}
\end{figure}

Figure \ref{pic:resolution} shows the midplane profiles of several key quantities zoomed in on the rim transition. The crosses represent grid points at the resolution used in the cases presented in this work. They form continuous, smooth transitions, which are resolved over several grid cells. No jumps or oscillations are visible, therefore the quantities are well resolved. Especially the critical region, where the optical depth $\tau$ increases from 0.1 to 10, is resolved over 103 grid points in the midplane. Additionally to ensure that the numerical resolution is sufficient a fiducial run at double resolution was performed.

\section{Results}
\label{sec:results}

This chapter presents the findings produced by introducing the new dust-to-gas ratio and the influence of diffusion and the temperature range, as well as the effect on the waves that were observed with the previous implementation of the dust-to-gas ratio. The model parameters used are listed in table \ref{tab1}.
These values are chosen because they are typical for Herbig Ae stars \citep{vdA}.
The surface density is determined using the assumption of steady state with a accretion rate of $\dot{M} = 10^{-8} M_\odot \textrm{yr}^{-1}$. Except in section \ref{sec:waves}, where the accretion rate is set higher at $\dot{M} = 10^{-7} M_\odot \textrm{yr}^{-1}$ to visualize the waves. The lower rate is chosen such, that no dust free hole forms inside the disk, whereas the higher rate includes that feature.
In \citep{Schobert19} it was shown, that for accretion rates above $\dot{M} = 10^{-8} M_\odot \textrm{yr}^{-1}$ the accretion heating can generate a gaseous inner hole in the dust distribution. Inside this cavity waves in the dust density formed and these will be scrutinized in section \ref{sec:waves}.

\subsection{Differences through the new dust-to-gas ratio}
\label{sec:comp}
 \begin{figure*}
	\centering
	\input{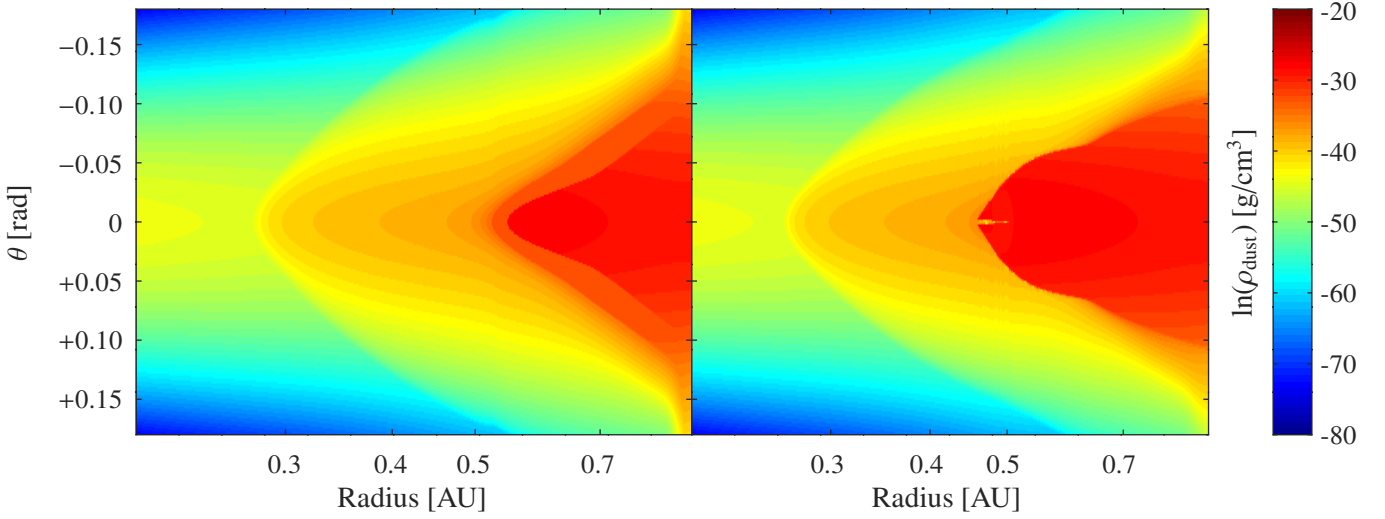}
	
	\caption{Comparison of old and new dust-to-gas ratio formula. Depicted is the natural logarithm of the dust density for both cases. The simulation on the left uses the old variant (eq. \eqref{d2g}), whereas the simulation whose result is shown on the right uses the new (eq. \eqref{eq:fd2gnew}). The $y$-axis is the polar angle in rad offset by $\pi/2$ and the $x$-axis the radial distance in AU.}%
	\label{pic:comp}
\end{figure*}

In fig. \ref{pic:comp} the dust density for the old (eq. \eqref{d2g}) and new (eq. \eqref{eq:fd2gnew}) dust-to-gas ratio are compared.  
The new dust-to-gas ratio formula changes the configuration of the inner region, specifically at the transition between vaporized and condensed dust. As can be seen in fig. \ref{pic:comp}, the transition radius lies further towards the star, from 0.55 AU to 0.45 AU. Also the condensed area is wider in $\theta$-direction at the transition radius and therefore has a steeper angle towards the stellar radiation.

The regions inwards of the transition radius are unaltered between the two versions, the halo forms in a identical shape and at the same distance to the star. Past 0.8 AU the models also agree with each other.

At the transition point in the midplane in the right panel in fig. \ref{pic:comp} there is a small line of condensed dust inside the disk. Because the stellar radiation is impinging on the disk directly perpendicular in the midplane, this point is prone to heat up and condense. In a physical disk with gas turbulence the dust above and below would be diffused into that line and close this gap. In the following section this effect is introduced and even the smallest amount of dust diffusion tested already effectively mitigates this artifact.

The old dust-to gas ratio kept the amount of dust at the transition artificially small, using $f_{\Delta \tau}$  so that the absorption of the radiation at the rim can be resolved \citep{FL16}. If the amount of dust is artificially diminished the evaporation front forms further away from the star. Since $f_{\Delta \tau}$ is a function of the density at each point, the diminishing effect is increased for smaller densities or equivalently further away from the midplane. Removing this artificial factor will therefore move the evaporation further towards the star and more so further above and below the midplane. These two effects are exactly what we see in fig. \ref{pic:comp}. Raising the resolution will increase $f_{\Delta \tau}$ and remedy these effects, beginning in the midplane, as is shown in the appendix of \citep{FL16}. To completely remedy the effect of $f_{\Delta \tau}$ far above and below the midplane, resolutions would be necessary that are not obtainable. The resolution would need to be increased in both dimensions by at least a factor 100. Therefore it is desired to remove $f_{\Delta \tau}$ from the equation completely. Doing so, however, creates artifacts connected with a poorly resolved radiation absorption, as can be seen in fig. \ref{pic:comp}. To remove these artifacts, dust diffusion is introduced as it provides a physically motivated smoothing effect allowing the radiation absorption to be resolved.  This is already achieved for even for the smallest amount of diffusion tested. Increasing the effect of diffusion to a reasonably motivated amount will change the shape of the rim to an even greater degree as is dicussed in the following section.

\subsection{Influence of the diffusion}
\label{sec:diffTime}
   \begin{figure*}
   \centering
   \input{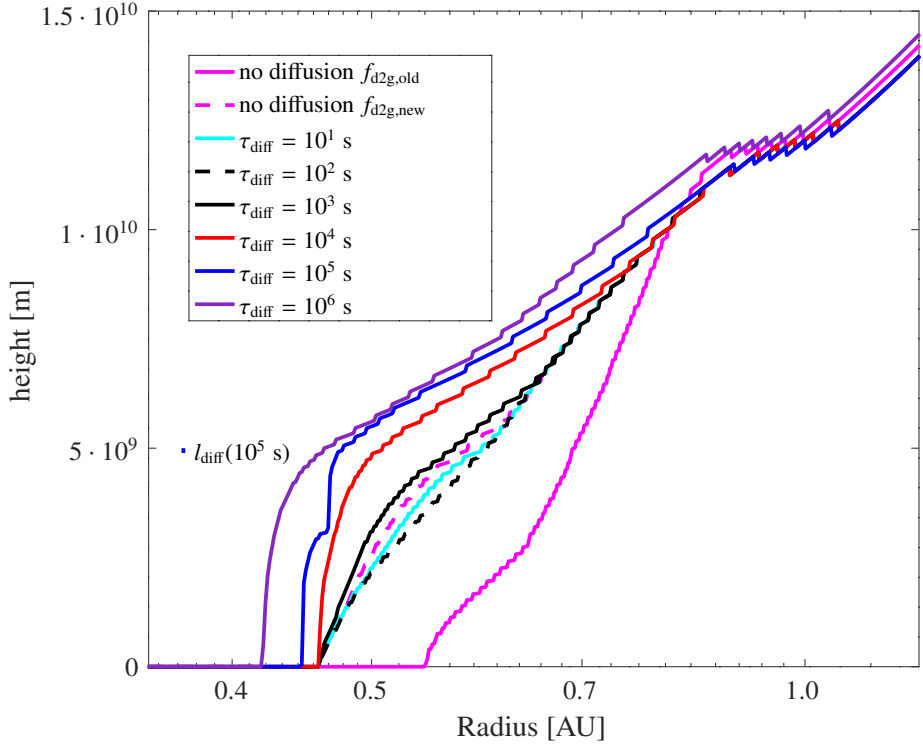}
   \caption{Comparison of eight radial disk height profiles at different diffusion time scales. Depicted is the height vertically above the midplane where the optical depth measured vertically towards the midplane from the box boundary reaches unity. The two magenta lines are cases with no dust diffusion or $\tau_\textrm{diff}=0$ for the old (solid) and new (dashed) versions of $f_\textrm{d2g}$ as described in section \ref{sec:d2g}. The six other cases cover the range of diffusion times from 10 seconds to $10^6$ seconds. The small blue bar to the left represents one diffusion length in the vertical direction for $\tau_\textrm{diff}=10^5$ s. It is significantly smaller than the change in shape of the corresponding blue profile.}%
   \label{pic:diffTime}
    \end{figure*}
This section elaborates on the influence of the diffusion. Eight runs with different diffusion times $\tau_\textrm{diff}$ were performed (see fig. \ref{pic:diffTime}). The times were chosen such, that they encompass the possible range of formation times of dust particles as explained in sec. \ref{sec:diff}. The profiles show where light falling vertically onto the disk reaches an optical depth of unity, which is used as a measure for the height of the disk. The first difference to note is between the two magenta lines, which denote the different ways of calculating the dust-to-gas ratio. With the new formula, eq. \eqref{eq:fd2gnew} (dashed line), the inner regions of the rim move closer to the star and are more rounded, while the outer regions stay almost unchanged. This is discussed in more detail in the previous sec. \ref{sec:comp}.
   
All the profiles with diffusion times given in this section use the new dust-to-gas formula. Comparing the profiles for no diffusion up to a diffusion time $\tau_\textrm{diff}=10^3$ shows little difference in the profile. The starting point in the midplane remains the same, then from $0.5\,$AU to $0.6\,$AU they deviate slightly from another and past that they align again. At first the height slightly decreases with higher diffusion times up to $\tau_\textrm{diff}=10^2$ but this trend stops with $\tau_\textrm{diff}=10^3$, when the height of the disk begins to increase with diffusion time. Overall the deviances for these four runs are small compared to the ones with higher diffusion times.

Comparing the profiles for diffusion times $\tau_\textrm{diff}=10^3$ up to $\tau_\textrm{diff}=10^6$ shows a stronger influence of diffusion on the profiles. First the starting point of the profile is closer to the star for higher diffusion times moving the inner rim inwards and second the height of the inner regions increases significantly for the closer regions (0.4 AU to 0.5 AU) and moderately for the mid region (0.5 AU to 0.8 AU). The outer regions again align with another. This leads to a change in the configuration of the disk. While disks at $\tau_\textrm{diff}=10^2$ have a narrow rounded off inner region, disks with diffusion times of $\tau_\textrm{diff}=10^4$ and higher display an enlarged surface of the inner region with a higher grazing angle. Although a higher grazing angle means a larger area facing the star and therefore more absorption of stellar irradiation, this is mitigated by a more diffused and therefore gradual dust transition and a new equilibrium is found.

Interesting to note is the discrepancy between the expected diffusion length $l_\textrm{diff}$ and the actual scale of the effect.
The typical diffusion length is 
\begin{equation}
l_\textrm{diff} = \sqrt{\tau D_\textrm{dust}}
\end{equation}
and can be used to compare the effect of dust diffusion on the rim shape with what could be expected. For typical values used in this paper $\tau = 10^5$ seconds and $D_\textrm{dust} = 1.27 \cdot 10^{11} \, \textrm{m}^2 \textrm{s}^{-1}$ calculated as described in section \ref{sec:diff}, the diffusion length is $l_\textrm{diff} = 1.1 \cdot 10^{8}\,\text{m}= 7.5 \cdot 10^{-4}\,\text{AU}$. The displacement of the $\tau = 1$ surface as seen in figure \ref{pic:diffTime} is a order of magnitude bigger than one diffusion length. This suggests a mechanism that reinforces the effect of diffusion on the rim. Potentially the equilibrium disk height with diffusion is additionally increased because the radiation from the star is absorbed over a greater distance and more gradually which allows for a steeper grazing angle in return. The smoother the transition into dust happens, the steeper the grazing angle can be.
This is one explanation for the strong effect of diffusion on the rim shape, that is elaborated on in sec. \ref{sec:disShape}.

  
\begin{table}
	\caption{\label{tab1}General model setup parameters.}

	\begin{tabular}{@{}llll}
			\hline
			\noalign{\smallskip}
			Parameter&Value\\
			\noalign{\smallskip}
			\hline
			\noalign{\smallskip}
			$N_r \times N_\theta$&$2560 \times 257$\\
			$[r_{\mathrm{min}},\,r_{\mathrm{max}}]$&$[0.2\,\textrm{AU},\,4\,\textrm{AU}]$\\
			$[\theta_{\mathrm{min}},\,\theta_{\mathrm{max}}]$&$[\pi/2-0.18, \,\pi/2+0.18]$\\
			Stellar parameter&$T_* = 10 000$ K, $R_*=2.5 R_\odot$, $M_*=2.5M_\odot$\\
			Dust-to-gas ratio& $f_0=0.01$\\
			\noalign{\smallskip}
			\hline
		\end{tabular}

\end{table}

\subsection{Influence of the temperature range}
\label{sec:delT}

\begin{figure}[htb]
	\centering
	\input{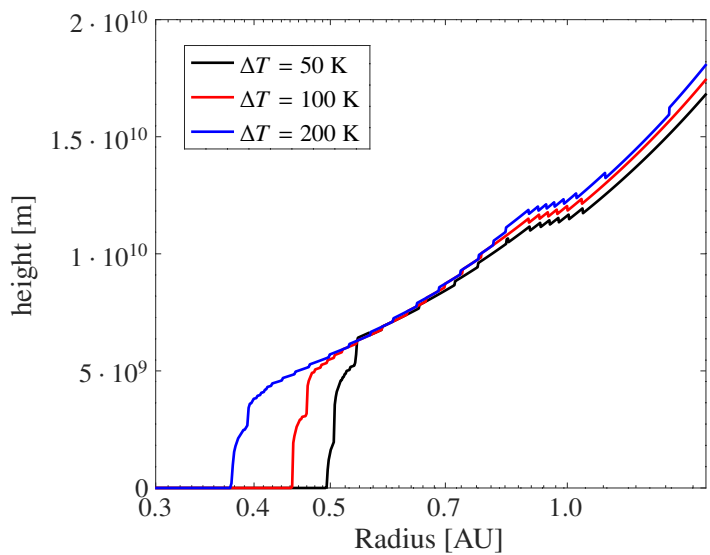}
	\caption{Comparison of three radial disk height profiles for different evaporation temperature ranges. Depicted is the height vertically above the midplane where the optical depth measured vertically towards the midplane from the box boundary reaches unity. The the black line is a case with $\Delta T = 50\,$K, the red line represents $\Delta T = 100\,$K and blue $\Delta T = 200\,$K.}%
	\label{pic:tempRange}
\end{figure}
This section describes the impact of the temperature range $\Delta T_\textrm{dust}$ over which evaporation occurs. The parameter is introduced in equation \eqref{eq:fd2gnew} and its default value is {100\,K}. Figure \ref{pic:tempRange} shows a comparison between three otherwise identical cases where $\Delta T_\textrm{dust}$ is {50\,K} (black line), {100\,K} (red line) and {200\,K} (blue line) respectively. The magnitude of the temperature range $\Delta T_\textrm{dust}$ can be observed to have little impact on the configuration of the rim. All three cases form a flattened structure with a nearly vertical cusp. However the range does affect the radius at which the condensation front facing the star forms. In the midplane the front moves inward from {0.49\,AU}, over {0.45\,AU} to {0.37\,AU}. If dust can condense at a higher temperature due to the larger range, the optical depth measured radially outward from the star grows faster. This means the region behind the point, where the first dust particles start to appear ($T \approx T_\text{ev}- 0.5 \Delta T$), receives less irradiation and the condensation front ($T = T_\text{ev}$) moves inward.

Different temperature ranges simulate different dust compositions of the disk. Depending on the size of the grains and their chemical components they evaporate at different temperatures. Since the configuration of the rim is not greatly affected by the temperature range, the findings of this paper can be applied to a wide variety of disks.

\subsection{Waves}
\label{sec:waves}

\begin{figure*}
	\centering
	\input{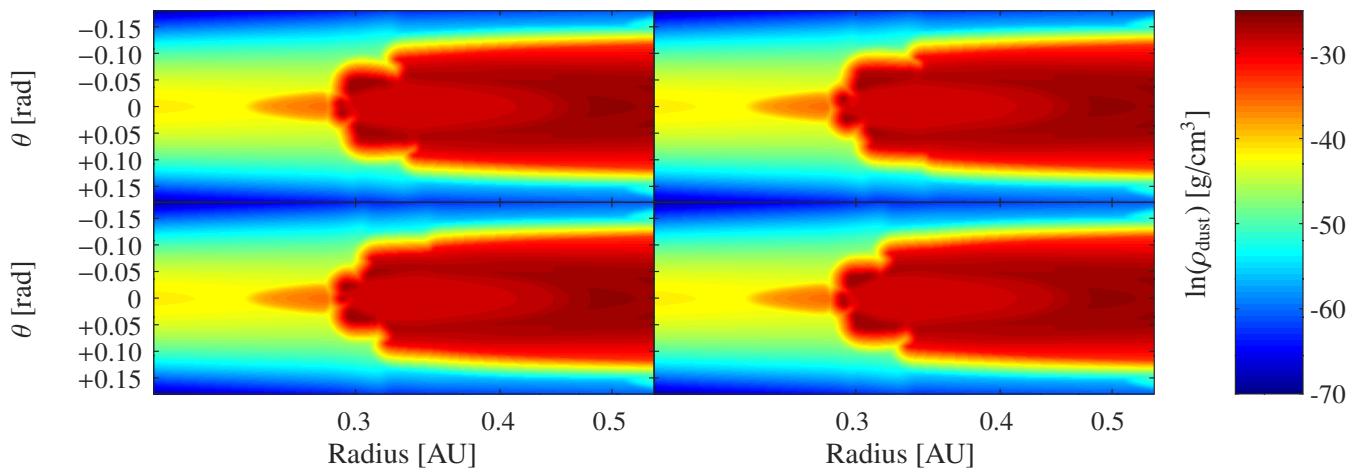}
	\caption{Evolution of one wave cycle. Depicted is the natural logarithm of the dust density at four points in time. The $y$-axis is the polar angle in rad offset by $\pi/2$ and the $x$-axis the radial distance in AU.}%
	\label{pic:waves}
\end{figure*}
%


In this section a case with an increased accretion rate $\dot{M} = 10^{-7} M_\odot \textrm{yr}^{-1}$ is explored. The dust diffusion time $\tau_\textrm{diff}$ is set to $10^5$ sec and the temperature range is 100 K.

At these parameters an inner hole forms inside the disk because the heat produced through viscous heating by accretion is sufficiently trapped inside the disk and the midplane heats above the evaporation point as was already explained in \citep{Schobert19}. Inside this inner hole waves in the dust density were travelling through the simulation domain away from the star. However there are two differences to the case in \citep{Schobert19}: first the dust walls caused by small temperature deviations that were observed in the previous work are no longer present. This suggests that they were an artifact of the discontinuous dust-to-gas ratio description, as was presumed in that paper. Since the dust-to-gas ratio is less sensitive around the evaporation temperature a continuous temperature region is present inside the hole with a smooth dust density.

The second effect is that waves in the dust density traverse along the outer perimeter of the disk away from the star. They originate at the closest distance of the disk to the star and then progress away from the former while continuously losing intensity. Eventually the fade into the body of the disk. One such cycle can be seen in fig. \ref{pic:waves}. The progression is left to right, then top to bottom with $3 \cdot 10^4$ seconds between each panel.

The first step is the formation of a new dust front at transition radius, which is at approximately 0.3 AU for this case. This front then gradually grows bigger and starts to move radially outwards. Once the front is big enough  it splits along the midplane and two separate dust density maxima move along the outer edge of the disk. Along their way they lose intensity and morph into the disk before they reach 0.4 AU. After the dust front has split it leaves the front of the disk open for the next dust front to form and the cycle continues. Fourier analysis of the temperature over time at a fixed position in the midplane shows an average dominant wave frequency at $2.19 \cdot 10^{-6}$ Hz or a period of 5.3 days. Coincidentally fluctuations of similar period have been observed in protostellar disks \citep{Fla14}. For the shown case the effect of these density fluctuations on synthetic SEDs of the disk has been found to be negligible using RADMC3D by \cite{2012ascl.soft02015D}, but the intensity of these waves depends on the diffusion time chosen, as well as the surface density of the disk (which is here determined through the accretion rate). These waves are found consistently for every parameter set, but are more pronounced for longer diffusion times as well as higher surface densities.

\section{Discussion}
\label{sec:discussion}

This chapter discusses the results in comparison to another theoretical model and quantifies the effects on rim shape and grazing angle. Synthetic images are shown that represent what different levels of diffusion would look like observationally.

\subsection{Theoretical comparison}

The cylindrical shape of the rim towards the star is reminiscent of a model previously proposed by \cite{DDN}. This section will compare this model to the findings of this paper, specifically the rim radius and rim height.

The rim position was determined in \citep{DDN} as the radius, where radiative equilibrium is reached at the evaporation temperature:

\begin{equation}
	R_\textrm{rim} = \left(\frac{L_*}{4 \pi T_\textrm{rim}^4 \sigma}\right)^{1/2}\left(1+ \frac{H_\textrm{rim}}{R_\textrm{rim}}\right),
	\label{eq:rimRad}
\end{equation}
where $L_*$ is stellar luminosity, $T_\textrm{rim}$ the evaporation temperature at the rim, $\sigma$ the Stefan-Boltzmann constant and $H_\textrm{rim}= \chi_\textrm{rim} h_\textrm{rim}$ the hight of the disk at the rim, which is a $\chi_\textrm{rim}$ times the pressure scale height at that location. The second factor of this equation takes the radiation into account that is reemitted from the opposing disk area.

To calculate $\chi_\textrm{rim}$ a vertical Gaussian density profile is assumed, as well as a linear behavior of $H(R)/R$ with a slope of $-1/8$ (see section A3 in \citep{DDN}). There the surface height is defined to be the height to which the optical depth $\tau$ of the rim on a radially outward directed ray is greater than 1:
\begin{equation}
\begin{aligned}
	\tau(H_\textrm{rim}) = \int_{R_\textrm{rim}}^{\infty} \rho(R, H_\textrm{rim}) \kappa_\textrm{dust}(T_*) f_\textrm{d2g} dR = 1 \\
	\textrm{erf}\left(\frac{\chi_\textrm{rim}^{\textrm{theo}}}{\sqrt{2}}\right) = 1 - \frac{1}{4 \Sigma(R_\textrm{rim}) \kappa_\textrm{dust}(T_*) f_\textrm{d2g} },
\end{aligned}
\end{equation}
here erf(x) denotes the error-function and $\Sigma$ is the gas surface density of the disk.

While in \citep{DDN} $\Sigma$ is chosen to be constant, this work has a radially depended surface density. $\Sigma(R_\textrm{rim})$ therefore needs to be chosen self-consistently to calculate $\chi$ which in turn is needed to iteratively find $R_\textrm{rim}$ with eq. (\ref{eq:rimRad}).

\begin{figure}[htb]
	\centering
	\input{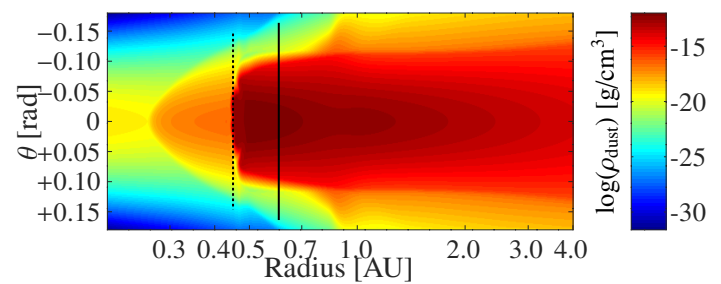}
	\caption{Comparison between analytical prediction and simulation. The dashed line shows $R_\textrm{rim}$ and the disk height calculated from the simulation and the solid line calculated as in \citep{DDN}. Depicted is the natural logarithm of the dust density for the case with $\Delta T = 100\,$K.}%
	\label{pic:dullComp}
\end{figure}

This yields $\chi_\textrm{rim}^{\textrm{theo}}= 3.69$ and $R_\textrm{rim}^{\textrm{theo}} = 0.64$ AU. However, since the simulation does not take into account the reemitted radiation at of the opposing dust rim, it is advantageous for the comparison to neglect the second factor of eq. (\ref{eq:rimRad}), which yields $\chi_\textrm{rim}^{\textrm{theo}}= 3.69$ and $R_\textrm{rim}^{\textrm{theo}} = 0.60$ AU. The height ratio $\chi$ does not change within two decimals, because the change in surface density is only small and the radius moves slightly inwards as expected if less radiation falls onto the rim. These latter two values are illustrated in fig. \ref{pic:dullComp} by the solid line.

For comparison the same quantities can be calculated from the simulation data, where $R_\textrm{rim}^{\textrm{sim}}$ is the radius, where the optical depth in the midplane measured from the star reaches 1 and $\chi_\textrm{rim}^{\textrm{sim}}$ is found through the height, where the optical depth at the boundary facing away from the star minus the optical depth at the rim radius reaches 1.

This yields $\chi_\textrm{rim}^{\textrm{sim}}= 4.13$ and $R_\textrm{rim}^{\textrm{sim}} = 0.45$ AU and is illustrated in fig. \ref{pic:dullComp} by the dotted line. The discrepancy in radius can be explained by the difference in slowly increasing the dust-to-gas ratio with the radius as in the simulation and switching from no dust being present at all to the maximum dust-to-gas ratio in the theoretical model. As more stellar radiation is absorbed closer to the star the equilibrium point moves inward as well.  The same effect was observed in sec. \ref{sec:comp}, when the discontinuity in the dust-to-gas model was removed the rim radius moved radially inward. The difference in $\chi$ is largely due to the change of pressure scale height with radius, it is conducive to also compare the radians covered by the different heights: $\theta_\textrm{sim}= 0.29$ and $\theta_\textrm{theo}= 0.33$ and these are in reasonably good agreement with each other as can be seen in fig. \ref{pic:dullComp}. This comparison shows that while an accurate assessment of course demands a numerical simulation, the analytic estimates provide a good approximation.

Another comparison can be drawn to observational data from \citep{Laza17}, where the height of the inner rim was estimated by considering the fraction of the stellar luminosity that is reprocessed to near-infrared radiation (NIR). That approach leads to $z/R \approx 0.2$, which exceeds predictions of previous hydrostatic models by a factor of two \citep{Vin07, Mul12}. 
However, it only exceeds the height of the inner rim, that is discussed in this section $z/R \approx 0.5 \, \theta_\textrm{sim} = 0.145$ by $38 \%$. This discrepancy can be in part due to the different methods of evaluating the rim height and the specific choice of parameters for the case presented in this section. Nevertheless, incorporating dust diffusion into hydrostatic models is a possible solution to the mismatch of disk heights in previous simulations and the disk heights observed in \citep{Laza17}.

\subsection{Rim shape}
\label{sec:disShape}

\begin{table*}
	\caption{Properties of different models discussed in this paper}
	\label{tab:props}
	
	\begin{tabular}{@{}llllllllllllll}
		\hline
		\noalign{\smallskip}
		Series&Figure&$\tau_\textrm{diff}$&$\Delta T_\textrm{dust}$&$R_1/$AU&$R_2/$AU&$\chi_2$&$z_2/R_2$&$R_3/$AU&$\chi_3$&$z_3/R_3$&$\beta_1$&$\beta_2$&$\beta_\textrm{total}$\\
		\noalign{\smallskip}
		\hline
		\noalign{\smallskip}
		Diffusion&2&0 sec&100 K&0.46&0.50&2.03&0.034&0.54&3.49&0.051&23.9$^\circ$&14.2$^\circ$&19.0$^\circ$\\
		time&&10 sec&100 K&0.46&0.50&1.80&0.030&0.54&3.19&0.047&21.3$^\circ$&13.6$^\circ$&17.3$^\circ$\\
		&&10$^2$ sec&100 K&0.46&0.49&1.31&0.022&0.54&2.90&0.043&22.7$^\circ$&12.4$^\circ$&15.6$^\circ$\\
		&&10$^3$ sec&100 K&0.46&0.46&0.55&0.009&0.53&3.61&0.053&43.6$^\circ$&19.9$^\circ$&21.8$^\circ$\\
		&&10$^4$ sec&100 K&0.46&0.47&2.09&0.034&0.52&4.72&0.068&69.6$^\circ$&19.9$^\circ$&30.7$^\circ$\\
		&&10$^5$ sec&100 K&0.45&0.45&1.78&0.029&0.53&5.27&0.076&83.4$^\circ$&19.5$^\circ$&27.0$^\circ$\\
		&&10$^6$ sec&100 K&0.42&0.42&0.48&0.008&0.78&5.21&0.092&90.0$^\circ$&10.8$^\circ$&11.3$^\circ$\\
		Temperature&3&10$^5$ sec&50 K&0.49&0.50&0.65&0.011&0.56&5.20&0.078&74.1$^\circ$&30.5$^\circ$&33.2$^\circ$\\
		range&&10$^5$ sec&200 K&0.37&0.38&1.18&0.018&0.78&4.90&0.086&77.3$^\circ$&8.6$^\circ$&9.4$^\circ$\\
		\noalign{\smallskip}
		\hline
	\end{tabular}
	
\end{table*}

To characterize the rim shape it is helpful to divide the rim into two sections. The first one extends from $R_1$, the radius where the vertical optical depth in the in the near infra-red becomes non zero and ends at $R_2$, the radius where the radial optical depth for the starlight is one. This region encompasses the very beginning of the disk and is near vertical for higher diffusion cases. The angle $\beta_1$ is calculated from the disk height at $R_2$ and the difference of these two radii.

The second section lies beyond $R_2$ and reaches until $R_3$, the radius where the midplane temperature falls below 1000 K. Corresponding to this section an angle $\beta_2$ can be calculated, which is the average disk angle between $R_2$ and $R_3$. The angle $\beta_\textrm{total}$ is the average angle of the disk from $R_1$ to $R_3$
The values $\chi_2$ and $\chi_3$ are the ratios of disk height, calculated with the vertical NIR optical depth, and pressure scale height at $R_2$ and $R_3$.

In table \ref{tab:props} the values for $\beta_1$ suggest that the diffusion does not significantly influence the angle of the first section up to a diffusion time of $10^2$ seconds. For higher diffusion times the angle rapidly steepens until it is fully vertical for the final value. The inner radius $R_1$ stays identical within two decimals up to a diffusion time of $10^5$ seconds then it starts to slowly decrease with it. Similarly $R_2$ moves inward with higher diffusion times and the distance between theses two radii decreases. The effect of the diffusion time on $R_3$ is small except for the two cases with the largest dust transition ranges $\tau_\textrm{diff} = 10^6$ sec and $\Delta T = 200$ K, where the midplane temperature dropped further away from the star, because the heating through irradiation could reach further inside the disk.

None of the cases tested displayed a shadow cast on the outer region of the disk. The grazing angle tends to zero before the disk starts to flare again further out. The change in stellar flux absorbed per unit distance monotonously decreases with the radius. This is in contrast to the theoretical model from the previous section. Heating by accretion keeps the disk hot enough to not collapse into a shadowed region.

The change of the rim shape due to diffusion of dust is quite dramatic compared to expectation informed by the diffusion length. As mentioned in section \ref{sec:diffTime} the displacement of the transition in vertical direction is 2 orders of magnitude larger then predicted. Because the transition between evaporated and condensed dust is such a delicate problem with repercussions for all areas behind it, already small differences can inform significant change in the overall configuration. Therefore dust diffusion plays an important role in the formation of the inner rim.

The introduction of dust diffusion allows dust to exist for a short period of time outside of areas where the model previously (without dust diffusion) predicted it. This newly displaced dust will have a cooling effect on everything in its shadow. More dust particles can form because of the lower temperature in these areas, enhancing the dust-to-gas ratio. Thus shifting the position of the dust-to-gas transition. Dust will be displaced by diffusion once again until equilibrium is reached. The shape of the rim is substantially altered through dust diffusion, although the diffusion length is small compared to this change. But the diffusion length is not small compared to all important scale lengths. The critical distance over which the dust-to-gas transition occurs is comparable to the diffusion length.

\subsection{Synthetic images}

\begin{figure}[htb]
	\centering
	\input{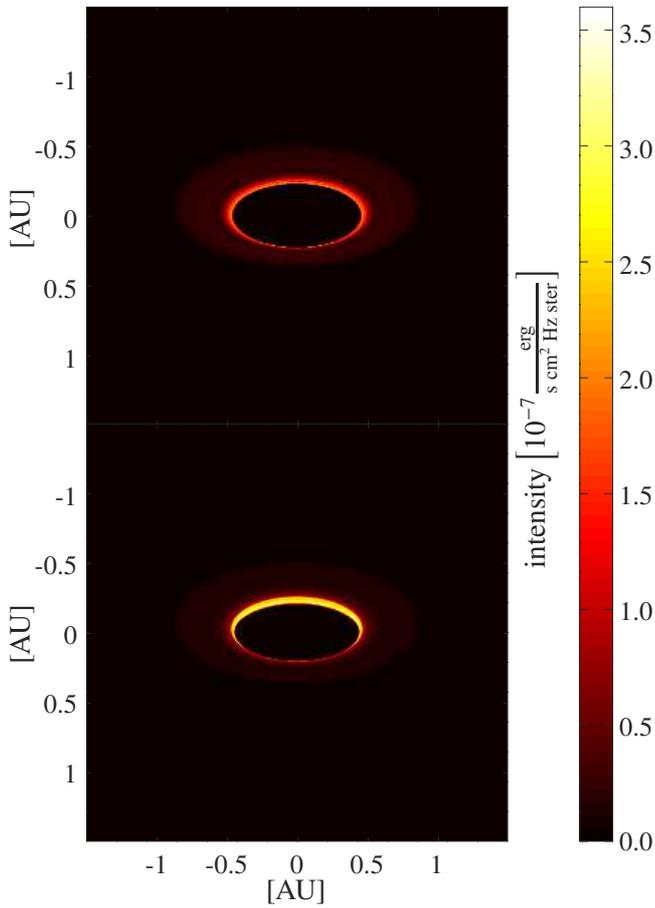}

	\vspace{-2cm}
	\caption{Synthetic images of the disk, viewed $60^\circ$ from face-on at $2\, \mu$m wavelength. The intensity maps correspond to no diffusion (top) and $\tau_\textrm{diff}= 10^5$ sec (bottom). }%
	\label{pic:imageComp}
\end{figure}

To provide a realistic view of the rim synthetic images were created using the RADMC3D code from \citep{2012ascl.soft02015D}. Fig. \ref{pic:imageComp} shows the intensity at a wavelength of 2 $\mu$m for a disk with a diffusion time $\tau_\textrm{diff}=0$s or diffusion turned off (upper panel) and a disk with $\tau_\textrm{diff}=10^5$s representing moderate diffusion (lower panel). The disks are viewed from above at an angle of $60^\circ$ towards face-on orientation, so the rim can be seen. The luminosity from the star itself is suppressed and the flux conservation is of second order.

The case without diffusion shows a thinner and rounder rim compared to the one with diffusion in agreement with fig. \ref{pic:diffTime}. The peak luminosity is reached at the rim itself in both cases, however the maximum is $2.1 \cdot 10^{-7} \textrm{erg}[\textrm{s cm$^2$ Hz ster}]^{-1}$ without diffusion and $3.5 \cdot 10^{-7} \textrm{erg}[\textrm{s cm$^2$ Hz ster}]^{-1}$ with diffusion. This is a 67\% increase of luminosity in the rim region.

The total luminosity is $3.12 \cdot 10^{-4} \textrm{erg}[\textrm{s cm$^2$ Hz ster}]^{-1}$ without diffusion and $3.31 \cdot 10^{-4} \textrm{erg}[\textrm{s cm$^2$ Hz ster}]^{-1}$ with diffusion. This is a 6\% increase of luminosity in the region of the NIR bump.

\section{Conclusions}
\label{sec:concl}
This paper presents a two-dimensional model for protoplanetary disks that synchronizes the dust description over all temperature ranges and consistently includes dust diffusion. The model builds upon earlier work \citep{FL16} and \citep{Schobert19} leading to qualitatively new results:

   \begin{enumerate}
   	  \item Dust diffusion has a far larger impact than a simple estimate of the diffusion length suggests. A feedback leads to a more gradual dust transition, which absorbs the stellar radiation along a longer path. This allows for a steeper grazing angle of the inner rim and a wider disk height.
      \item Varying the temperature range moves the inner radius, while the general structure of the rim is unchanged. This means the model is robust for different possible dust compositions.
      \item The intensity emitted by the disk at a wavelength of 2 $\mu$m is a function of the diffusion time. It increases significantly at the apex of the rim and slightly when averaged over the whole disk.
      \item No more waves are observed inside the gaseous hole as found in \cite{Schobert19}, those were an artifact of the discontinuous dust description.
      \item Waves in dust density propagate along the outer perimeter of the disk for higher accretion rates and lower dust diffusion times. These display a period of 5.3 days.
   \end{enumerate}

The code used to produce these results can be found at \mbox{bitbucket.org/astro\_bayreuth/radiation\_code}.

\bibliographystyle{aa} 		
\bibliography{references} 	

\begin{thebibliography}{20}
\expandafter\ifx\csname natexlab\endcsname\relax\def\natexlab#1{#1}\fi

\bibitem[{Brauer {et~al.}(2008)Brauer, Dullemond, \& Henning}]{DULL08}
Brauer, F., Dullemond, C.~P., \& Henning, T. 2008, A\&A, 480, 859

\bibitem[{Dullemond {et~al.}(2001)Dullemond, Dominik, \& Natta}]{DDN}
Dullemond, C.~P., Dominik, C., \& Natta, A. 2001, ApJ, 560, 957

\bibitem[{{Dullemond} {et~al.}(2012){Dullemond}, {Juhasz}, {Pohl}, {Sereshti},
  {Shetty}, {Peters}, {Commercon}, \& {Flock}}]{2012ascl.soft02015D}
{Dullemond}, C.~P., {Juhasz}, A., {Pohl}, A., {et~al.} 2012, {RADMC-3D: A
  multi-purpose radiative transfer tool}

\bibitem[{Duschl {et~al.}(1996)Duschl, Gail, \& Tscharnuter}]{Duschl96}
Duschl, W.~J., Gail, H.-P., \& Tscharnuter, W.~M. 1996, A \& A, 312, 624

\bibitem[{Flaherty {et~al.}(2014)}]{Fla14}
Flaherty, K.~M. {et~al.} 2014, ApJ, 793, 2

\bibitem[{Flock {et~al.}(2016)Flock, Fromang, Turner, \& Benisty}]{FL16}
Flock, M., Fromang, S., Turner, N.~J., \& Benisty, M. 2016, ApJ, 827, 144

\bibitem[{Isella \& Natta(2005)}]{IN}
Isella, A. \& Natta, A. 2005, A \& A, 438, 899

\bibitem[{Lazareff {et~al.}(2017)Lazareff, Berger, \& Kluska}]{Laza17}
Lazareff, B., Berger, J.-P., \& Kluska, J. a.~o. 2017, A\&A, 599, A85

\bibitem[{{Lenzuni} {et~al.}(1995){Lenzuni}, {Gail}, \& {Henning}}]{Lenz95}
{Lenzuni}, P., {Gail}, H.-P., \& {Henning}, T. 1995, \apj, 447, 848

\bibitem[{Li \& Draine(2001)}]{LiDrain}
Li, A. \& Draine, B.~T. 2001, ApJ, 554, 778

\bibitem[{Morfill(1988)}]{MOR88}
Morfill, G.~E. 1988, Icarus, 75, 371

\bibitem[{{Mulders, G. D.} \& {Dominik, C.}(2012)}]{Mul12}
{Mulders, G. D.} \& {Dominik, C.} 2012, A\&A, 539, A9

\bibitem[{Nakamura {et~al.}(2007)}]{NA07}
Nakamura, T.~M. {et~al.} 2007, METEORIT PLANET SCI, 42, 1249

\bibitem[{Schobert {et~al.}(2019)Schobert, Peeters, \& Rath}]{Schobert19}
Schobert, B.~N., Peeters, A.~G., \& Rath, F. 2019, ApJ, 881, 56

\bibitem[{Tachibana {et~al.}(2014)Tachibana, Takigawa, Miyake, Nagahara, \&
  Ozawa}]{Tach14}
Tachibana, S., Takigawa, A., Miyake, A., Nagahara, H., \& Ozawa, K. 2014, 45th
  Lunar and Planetary Science Conference, 1226

\bibitem[{Tachibana {et~al.}(2011)}]{TA11}
Tachibana, S. {et~al.} 2011, ApJ, 763, 16

\bibitem[{Tannirkulam {et~al.}(2007)Tannirkulam, Harries, \&
  Monnier}]{Tannir07}
Tannirkulam, A., Harries, T.~J., \& Monnier, J.~D. 2007, ApJ, 661, 374

\bibitem[{van~den Ancker {et~al.}(1998)van~den Ancker, de~Winter, \& Tjin
  A~Djie}]{vdA}
van~den Ancker, M.~E., de~Winter, D., \& Tjin A~Djie, H. R.~E. 1998, A \& A,
  330, 145

\bibitem[{Van~der Vorst(1992)}]{VDV}
Van~der Vorst, H.~A. 1992, SIAM J. Sci. Stat. Comput., 13, 631

\bibitem[{Vinkovi{\'{c}} \& Jurki{\'{c}}(2007)}]{Vin07}
Vinkovi{\'{c}}, D. \& Jurki{\'{c}}, T. 2007, The Astrophysical Journal, 658,
  462

\end{thebibliography}

\end{document}